\documentclass{article}

\usepackage{graphicx,float}
\usepackage{t1enc}
\usepackage{epsfig}
\usepackage[francais,]{babel}
\usepackage{amssymb,amsbsy,amsmath,amsfonts,amssymb,amscd}

\newtheorem{theoreme}{Th\'eor\`eme}[section]
\newtheorem{lemme}[theoreme]{Lemme}
\newtheorem{proposition}[theoreme]{Proposition}

\newtheorem{remarque}{\it Remarque}

\begin{document}
\title{\Huge{\bf Approche intrins\`{e}que des fluctuations quantiques en m\'{e}canique
stochastique}
\\ ~ \\ \Large{\bf{\em An intrinsic approach to the quantum fluctuations
in stochastic mechanics}}  \\ ~ \\}

\author{Michel FLIESS \\  ~ \\ {\small{ Projet ALIEN, INRIA Futurs}} \\
\small{ \& \'Equipe MAX, LIX (CNRS, UMR 7161)}
\\  \small{\'{E}cole polytechnique, 91128 Palaiseau, France} \\
\small{ E-mail: {\tt Michel.Fliess@polytechnique.edu}}}
\date{}


\maketitle {

\noindent{\bf Abstract}. This note is answering an old questioning
about the F\'{e}nyes-Nelson stochastic mechanics. The Brownian nature of
the quantum fluctuations, which are associated to this mechanics, is
deduced from Feynman's interpretation of the Heisenberg uncertainty
principle via infinitesimal random walks stemming from nonstandard
analysis. It is therefore no more necessary to combine those
fluctuations with a background field, which has never been well
understood.
\\

\noindent{\bf Key words}. Stochastic mechanics, quantum
fluctuations, nonstandard analysis, random walks, fractal curves,
stochastic differential equations.
\\
}

\section*{Abridged English version}
\noindent {\bf Introduction.} The background field of quantum
fluctuations in the F\'enyes-Nelson {\em stochastic mechanics}
\cite{fen,physrev,nelson-brown,nelson-quant} (see also
\cite{davidson,fri,fri-bis,nagasawa1,nagasawa2,oron,pavon,pen,smolin,smolin2})
has never received a clear-cut and well accepted justification. The
aim of this note is to provide a new intrinsic construction of those
fluctuations, which are of Brownian nature. Its main mathematical
tool is Robinson's nonstandard analysis \cite{robinson} and, more
precisely, the nonstandard presentation of stochastic differential
equations (see, e.g., \cite{al,benoit}).

\subsubsection*{Nontechnical presentation of the main ideas}
\noindent This summary is intended for readers who are not familiar
with nonstandard analysis. It shows that the Brownian fluctuations
are direct consequences of an infinitesimal time discretisation
combined with the Heisenberg uncertainty principle. As often in
nonstandard analysis (see, e.g., \cite{al,nelson-proba}), we replace
a continuous time interval by an infinite ``discrete'' set of
infinitely closed time instants. Substituting $m \frac{\Delta
x}{\Delta t}$ to $\Delta p$ in the well known expression of the
uncertainty principle, where $x$ is the position, $m$ the mass, $p$
the momentum, $h = 2 \pi \hbar$ the Planck constant, yields equation
(\ref{rhei}). We rewrite it by postulating that the quantity
(\ref{fond}), where $\delta t
> 0$ is a given infinitesimal, is limited and appreciable, i.e., it is
neither infinitely large nor infinitely small. Those computations
are stemming from Feynman's interpretation \cite{feynman0,feynman}
of the uncertainty principle (see, also, \cite{kr,not1}): The
``quantum trajectories'' are fractal curves, of Hausdorff dimension
$2$. ``Weak'' mathematical assumptions permit to derive the
infinitesimal difference equation (\ref{ma}). The lack of any
further physical assumption yields the equiprobability of $+1$ and
$- 1$. If $x$ is {\em Markov}, i.e., if $b$ and $\sigma$ are are
functions of $t$ and $x(t)$, and not of $\{ x(\tau) | 0 \leq \tau <
t \}$, the corresponding infinitesimal random walk is \og
equivalent\fg ~ to a stochastic differential equation in the usual
sense (see, e.g., \cite{al,benoit}).
\\

\noindent {\it Remark 1 More or less analogous random walks have
already been introduced in the literature (see, e.g.,
\cite{badiali,gud,kempe,ord1,ord2}), but in another context.}

\subsubsection*{Nonstandard analysis}
\noindent Replace the interval $[0, 1]$ by the set ${\mathfrak{Q}} =
\{k \delta t \mid 0 \leq k \leq N_q \}$, $\delta t = \frac{1}{N_q}$,
where $N_q$ is an unlimited integer. A function $x: {\mathfrak{Q}}
\rightarrow {\mathbb{R}}$ is said to verify the {\em Heisenberg
condition} if, and only if, for any $t \in {\mathfrak{Q}} \setminus
\{1\}$, equation (\ref{heis}) is satisfied, where
$\sigma^{\star}_{t} \in {\mathbb{R}}$, $\sigma^{\star}_{t}> 0$, is
limited and appreciable. The lack of any further physical assumption
leads us to postulate the following properties for $\varepsilon (t)$
in equation (\ref{heis-prob}): the random variables $\varepsilon
(t)$, $t \in {\mathfrak{Q}} \setminus \{1\}$, are independent;
$\mbox{\rm Pr} (\varepsilon (t) = +1) = \mbox{\rm Pr} (\varepsilon
(t) = -1) = \frac{1}{2}$. The next properties are ``natural'' for
the stochastic processes $\varepsilon$ and $x$ (see
\cite{nelson-proba} and \cite{benoit0,benoit}): the random variables
$x(0)$ and $\varepsilon (t)$, $t \in {\mathfrak{Q}}$, are
independent; $x(t)$, $t > 0$, is function of $\{\varepsilon (\tau)|
0 \leq \tau \leq t - \delta t \}$. Set $E^{\varepsilon}_{t}
(\bullet) = E_{t} (\bullet | \varepsilon (0), \varepsilon (\delta
t), \dots, \varepsilon (t - \delta t))$. Equation (\ref{qf}) yields
the decomposition (\ref{decomp}) \cite{nelson-proba}, where $\eta$
is a stochastic process, such that $E^{\varepsilon}_{t} (\eta (t)) =
0$, $E^{\varepsilon}_{t} ((\eta (t))^2) = 1$. Then, for all $t \in
{\mathfrak{Q}} \setminus \{1\}$, $\sigma^{\star}_{t} \simeq
s^{\star}_{t}$, $\varepsilon (t) = \eta (t)$.

We will not make any distinction between two equations of type
(\ref{decomp}) if the coefficients are limited and infinitely
closed. The process $x$ is said to be {\em Markov}, or to satisfy
the {\em Markov condition}, if, and only if, there exist $b \simeq
D^\varepsilon x (t)$, $\sigma \simeq s^{\star}_{t}$, such that they
are functions of $t$ et $x(t)$, and not of $\{x(\tau) |  0 \leq \tau
< t \}$. Assume that $x$ is Markov. Consider the infinitesimal
random walk (\ref{mad}). According to \cite{benoit0} (see, also,
\cite{benoit}) it defines a diffusion if, and only if, the following
conditions are satisfied (see \cite{al} for another approach): $b$
and $\sigma$ are of class $S^0$; the shadows $\tilde{b}$ and
$\tilde{\sigma}$ of $b$ et $\sigma$ are of class $C^\infty$; the
function $\tilde{\sigma}$ is always strictly positive; the random
variable $x(0)$ is almost surely limited.

\setcounter{section}{0} \selectlanguage{francais}
\section{Introduction}
\label{intro}

Invent\'{e}e par F\'enyes \cite{fen}, d\'{e}velopp\'{e}e et popularis\'{e}e par
Nelson \cite{physrev,nelson-brown,nelson-quant}, la {\em m\'{e}canique
stochastique} offre, ainsi que diverses variantes (renvoyons \`{a}
quelques travaux
\cite{davidson,fri,fri-bis,nagasawa1,nagasawa2,oron,pavon,pen,smolin,smolin2},
tous post\'{e}rieurs aux publications de Nelson), une alternative
passionnante aux fondements du quantique, d\'{e}j\`{a} recherch\'{e}e par
Schr\"{o}dinger \cite{schrodinger} (cf. \cite{zambrini}). Elle y
parvient par l'introduction d'un champ de fluctuations quantiques,
de nature brownienne, proche des bruits dans des domaines
d'ing\'{e}ni\'{e}rie, comme l'automatique et le signal. Nous justifions,
ici, ces fluctuations, sans la n\'{e}cessit\'{e} d'un tel champ, qui n'a pas
re\c{c}u de justifications convaincantes. Que l'on nous permette une
citation (\cite{nelson-quant}, p. 68): \og {\it Let me remark that I
have no evidence for the background field hypothesis (if I did, I
would gladly sacrifice an ox)} \fg! Voir les r\'{e}f\'{e}rences pr\'{e}c\'{e}dentes
pour un examen critique et des indications bibliographiques
suppl\'{e}mentaires. Cette note repose sur
\begin{itemize}
\item l'interpr\'{e}tation par Feynman \cite{feynman0,feynman} du
principe d'incertitude de Heisenberg;

\item l'{\em analyse non standard} de Robinson \cite{robinson}, d\'{e}j\`{a}
employ\'{e}e en quantique (voir, par exemple, \cite{al}, sa
bibliographie, et \cite{gud,not1}), et, plus pr\'{e}cis\'{e}ment,
\begin{itemize}
\item une
discr\'{e}tisation infinit\'{e}simale du temps,
\item  le calcul non standard des probabilit\'{e}s
de Nelson \cite{nelson-proba} et son extension aux \'{e}quations
diff\'{e}rentielles stochastiques \cite{benoit0,benoit}.
\end{itemize}
\end{itemize}
Terminons par une autre citation (\cite{nelson-quant}, p. 94) \`{a}
propos de l'exp\'{e}rience des fentes d'Young: \og {\em The particle
doesn't know whether the other slit is open, but the background
field does}\fg. Cet embarras se retrouve peu ou prou dans les
explications des paradoxes c\'{e}l\`{e}bres, fournies en
\cite{fri,fri-bis,nagasawa1,nagasawa2,nelson-quant,pen} et souvent
s\'{e}duisantes. Il semble, ici, dissip\'{e}.

\begin{remarque}Afin d'all\'{e}ger l'\'{e}criture, nous nous restreignons \`{a}
un espace de dimension $1$.
\end{remarque}

\begin{remarque}Alors que certains travaux autour de la m\'{e}canique
stochastique se rattachent \`{a} l'automatique et, plus pr\'{e}cis\'{e}ment, \`{a}
la commande optimale stochastique (cf. \cite{guerra}), notre point
de vue fait suite \`{a} une approche nouvelle \cite{fliess} de
l'estimation et de l'identification, en automatique et signal.
Rappelons les liens entre quantique actuel et certains aspects du
traitement du signal (cf. \cite{flandrin}).
\end{remarque}

\section{Pr\'{e}sentation g\'{e}n\'{e}rale}
Voici une synth\`{e}se des principales id\'{e}es directrices, sans
pr\'{e}tention de rigueur, mais, esp\'{e}rons-le, claire et transparente,
destin\'{e}e aux lecteurs peu au fait de l'analyse non standard:

\begin{enumerate}

\item On remplace, comme souvent en non-standard (cf. \cite{al,nelson-proba}), le temps
continu par un ensemble infini \og discret \fg ~ d'instants
infiniment proches.

\item
Les calculs suivants \'{e}manent d'une reformulation par Feynman
\cite{feynman0,feynman} du principe d'incertitude de Heisenberg
(voir, aussi, \cite{kr,not1}): la \og trajectoire quantique \fg ~
est une courbe continue, non d\'{e}rivable, c'est-\`{a}-dire fractale, de
dimension de Hausdorff $2$, traduisant la \og {\em Zitterbewegung}
\fg. Soient $x$ la position, $m$ la masse, $p$ la quantit\'{e} de
mouvement, $h = 2 \pi \hbar$ la constante de Planck. L'expression
famili\`{e}re $\Delta x \Delta p \gtrsim \hbar$ du principe
d'incertitude devient (cf. \cite{kr}, p. 85):
\begin{equation}\label{rhei}
\frac{(\Delta x)^2}{\Delta t} \gtrsim \frac{\hbar}{m}
\end{equation}
Soit $\delta t > 0$ infinit\'{e}simal donn\'{e}. On r\'{e}\'{e}crit (\ref{rhei}) en
postulant que
\begin{equation}\label{fond}\frac{ \left( x(t + \delta t) - x(t)
\right)^2}{\delta t}\end{equation} est limit\'{e} et appr\'{e}ciable,
c'est-\`{a}-dire ni infiniment grand ni infiniment petit.

\item On en d\'{e}duit, moyennant des hypoth\`{e}ses math\'{e}matiques \og faibles \fg,
l'\'{e}quation aux diff\'{e}rences infinit\'{e}simales
\begin{equation}\label{ma}
x(t + \delta t) = x(t) + b \delta t \pm \sigma \sqrt{\delta t}
\end{equation}

\item L'absence de toute hypoth\`{e}se physique suppl\'{e}mentaire conduit \`{a}
postuler l'\'{e}quiprobabilit\'{e} de $\pm 1$.

\item Si $x$ est {\em markovien},
c'est-\`{a}-dire si $b$ et $\sigma$ sont fonctions de $t$ et $x(t)$, et
non de $\{ x(\tau) | 0 \leq \tau < t \}$, cette marche al\'{e}atoire
infinit\'{e}simale \og \'{e}quivaut \fg ~ \`{a} une \'{e}quation diff\'{e}rentielle
stochastique au sens usuel (cf. \cite{al,benoit}).

\end{enumerate}

\begin{remarque}
Nos marches al\'{e}atoires, qui pourraient venir compl\'{e}ter les {\em
marches al\'{e}atoires quantiques} et leurs applications informatiques
(cf. \cite{kempe}), d\'{e}coulent de principes premiers. Les liens de
telles marches avec le quantique \'{e}taient d\'{e}j\`{a} connus (cf.
\cite{gud,ord1,ord2}). Voir, \`{a} propos de l'irr\'{e}versibilit\'{e}
thermodynamique, \cite{badiali} pour une condition voisine de
(\ref{rhei})-(\ref{fond}) dans un espace-temps discret.
\end{remarque}
\begin{remarque}
Ce travail est une autre manifestation des liens, d\'{e}j\`{a} \'{e}voqu\'{e}s en
maintes publications et soulign\'{e}s en \cite{not1}, entre m\'{e}canique
stochastique et approche fractale de la microphysique.
\end{remarque}

\begin{remarque}Le caract\`{e}re statistique du quantique, confirm\'{e}
par tant d'exp\'{e}riences, se fonde, comme on le sait, sur
l'interpr\'{e}tation, due \`{a} Born, de la fonction d'onde de Schr\"{o}dinger.
L'explication fournie ici est autre.
\end{remarque}

\section{Analyse non standard}
On r\'{e}dige selon la terminologie {\it IST} de
\cite{nelson-ist,nelson-proba}.

\subsection{Condition de Heisenberg}\label{cond}
Rempla\c{c}ons l'intervalle $[0, 1]$ par ${\mathfrak{Q}} = \{k \delta t
\mid 0 \leq k \leq N_q \}$, $\delta t = \frac{1}{N_q}$, o\`u $N_q$
est un entier illimit\'{e}. Appelons ${\mathfrak{Q}}$ et $\delta t$,
respectivement, l'{\em \'{e}chelle} et le {\em pas de temps quantiques}.
Une fonction $x: {\mathfrak{Q}} \rightarrow {\mathbb{R}}$ est dite
v\'{e}rifier la {\em condition de Heisenberg} si, et seulement si, pour
tout $t \in {\mathfrak{Q}} \setminus \{1\}$,

\begin{equation}
\label{heis} \frac{ \left( x(t + \delta t) - x(t) \right)^2}{\delta
t} \simeq ({\sigma}^{\star}_{t})^2 \end{equation} o\`{u}
${\sigma}^{\star}_{t} \in {\mathbb{R}}$, ${\sigma}^{\star}_{t} > 0$,
est limit\'{e} et appr\'{e}ciable. Il en d\'{e}coule

\begin{equation}
\label{heis-prob}  x(t + \delta t) - x(t) \simeq \varepsilon
(t)({\sigma}^{\star}_{t})\sqrt{\delta t} \quad \mbox{\rm o\`{u}} ~
\varepsilon (t) = \pm 1
\end{equation}

\subsection{\'Equiprobabilit\'{e} de $+1$ et $-1$}
Postulons les deux propri\'{e}t\'{e}s suivantes, qui d\'{e}coulent \og
naturellement \fg ~ de l'absence de toute hypoth\`{e}se physique
suppl\'{e}mentaire:
\begin{itemize}
\item les $\varepsilon (t)$, $t \in {\mathfrak{Q}} \setminus \{1\}$, sont des
variables al\'{e}atoires ind\'{e}pendantes,
\item $\mbox{\rm Pr}
(\varepsilon (t) = +1) = \mbox{\rm Pr} (\varepsilon (t) = -1) =
\frac{1}{2}$.
\end{itemize}

\subsection{Processus stochastiques}
Comme $\varepsilon$ peut \^etre vu comme un processus stochastique,
il est loisible de postuler les propri\'{e}t\'{e}s suivantes (cf.
\cite{nelson-proba} et \cite{benoit0,benoit}):
\begin{itemize} \item $x$ est un processus stochastique;
\item les variables al\'{e}atoires $x(0)$ et $\varepsilon (t)$, $t \in
{\mathfrak{Q}}$, sont ind\'{e}pendantes;
\item $x(t)$, $t > 0$, est fonction de $\{\varepsilon (\tau)| 0 \leq \tau \leq
t - \delta t \}$.
\end{itemize}
Notons $E^{\varepsilon}_{t} (\bullet)$ l'esp\'{e}rance conditionnelle du
processus $\bullet$ en l'instant $t \in {\mathfrak{Q}}$, sachant le
pass\'{e} du processus $\varepsilon$. Posons
\begin{equation}\label{qf}
\begin{array}{c}
D^\varepsilon x(t) = \frac{E^{\varepsilon}_{t} (x(t + \delta
t) - x(t))}{\delta t}  \\
s^{\star}_{t} \simeq \sqrt{\frac{(x(t + \delta t) - x(t) -
D^\varepsilon x(t) \delta t)^2}{\delta t}}
 \end{array}
\end{equation}
Supposons $D^\varepsilon x(t)$ limit\'{e}. On obtient la d\'{e}composition
\cite{nelson-proba} (voir, aussi, \cite{benoit0}):
\begin{equation}\label{decomp}
x (t + \delta t) = x(t) + (D^\varepsilon x(t)) \delta t +
(s^{\star}_{t}) \eta (t) \sqrt{\delta t}
\end{equation}
o\`u $\eta$ est un processus stochastique, tel que
$E^{\varepsilon}_{t} (\eta (t)) = 0$, $E^{\varepsilon}_{t} ((\eta
(t))^2) = 1$. La proposition suivante relie les quantit\'{e}s
$\sigma^{\star}_{t}$ et $\varepsilon (t)$ des paragraphes pr\'{e}c\'{e}dents
aux nouvelles:

\begin{proposition}
Pour tout $t \in {\mathfrak{Q}} \setminus \{1\}$, il est loisible de
poser $\sigma^{\star}_{t} \simeq s^{\star}_{t}$, $\varepsilon (t) =
\eta (t)$.
\end{proposition}

\subsection{Condition de Markov}
Soit
$$
x_i (t + \delta t) = x_i(t) + (D^\varepsilon x_i(t)) \delta t +
(s^{\star}_{i, t}) \varepsilon (t) \sqrt{\delta t}
$$
o\`u  $i = 1, 2$. La propri\'{e}t\'{e} suivante est facile:
\begin{lemme}
Si, pour tout $t \in {\mathfrak{Q}} \setminus \{1\}$, les quantit\'{e}s
infiniment proches $D^\varepsilon x_1(t) \simeq D^\varepsilon
x_2(t)$, $s^{\star}_{1, t} \simeq s^{\star}_{2, t}$, $x_1(0) \simeq
x_2 (0)$ sont limit\'{e}es en valeur absolue par $C \in \mathbb{R}$, la
diff\'{e}rence $x_1 (t) - x_2 (t)$ reste infinit\'{e}simale pour tout $t \in
\mathfrak{Q}$.
\end{lemme}
Nous ne distinguerons pas deux \'{e}quations de type (\ref{decomp}) avec
coefficients limit\'{e}s et infiniment proches. On dit que $x$ est {\em
markovien}, ou satisfait la {\em condition de Markov}, si, et
seulement si, il existe des quantit\'{e}s $b \simeq D^\varepsilon x
(t)$, $\sigma \simeq s^{\star}_{t}$ fonctions de $t$ et $x(t)$, et
non de $\{x(\tau) |  0 \leq \tau < t \}$.

\subsection{Diffusions}
Supposons, dor\'{e}navant, $x$ markovien et consid\'{e}rons la marche
al\'{e}atoire infinit\'{e}simale
\begin{equation}
\label{mad} x(t + \delta t) = x(t) + b(t, x(t)) \delta t + \sigma
(t, x(t)) \varepsilon (t) \sqrt{\delta t}
\end{equation} Renvoyons \`{a} \cite{nelson-proba} (voir, aussi,
\cite{benoit0,benoit}) pour la notion d'{\em \'{e}quivalence} de
processus stochastiques. Le repr\'{e}sentant (\ref{mad}) d\'{e}finit, selon
\cite{benoit0} (voir, aussi, \cite{benoit}), une diffusion si, et
seulement si, les conditions suivantes sont satisfaites:
\begin{itemize}
\item $b$ et $\sigma$ sont de classe $S^0$,

\item les ombres $\tilde{b}$ et $\tilde{\sigma}$ de
$b$ et $\sigma$ sont de classe $C^\infty$,
\item la fonction $\tilde{\sigma}$ est toujours strictement
positive,
\item la variable al\'{e}atoire $x(0)$ est presque s\^urement limit\'{e}e.
\end{itemize}

\begin{remarque}
Voir \cite{al} pour une autre approche des liens entre (\ref{mad})
et \'{e}quations diff\'{e}rentielles stochastiques, o\`u les conditions
requises pour $b$ et $\sigma$ sont sensiblement diff\'{e}rentes.
\end{remarque}

\vspace{0.1cm} {\small \noindent{\bf Remerciements}. L'auteur
exprime sa reconnaissace \`a E. Beno\^{\i}t (La Rochelle) pour lui avoir
communiqu\'{e} la r\'{e}f\'{e}rence \cite{benoit0}, \`{a} J.-M. L\'{e}vy-Leblond (Nice)
pour des remises en question salutaires, \`{a} P. Rouchon (Paris) pour
des discussions pr\'{e}cieuses, et \`{a} T. Sari (Mulhouse) pour des
conseils utiles.}

\end{document}